# Correlation between magnetism and the Verwey transition in magnetite

Karolina Podgórska[1], Mateusz A. Gala[1,2], Kamila Komędera[1,3], N. K. Chogondahalli Muniraju[4], Serena Nasrallah[2], Zbigniew Kąkol[1], Joseph Sabol[5], Christophe Marin[7], Adam Włodek[8], Andrzej Kozłowski[1], J. Emilio Lorenzo[9], Neven Barišić[2,10], Damian Rybicki[1], and Wojciech Tabiś[1*]

[1]*AGH University of Krakow, Faculty of Physics and Applied Computer Science, Aleja Mickiewicza 30, 30-059 Kraków, Poland*

[2]*Institute of Solid State Physics, TU Wien, 1040 Vienna, Austria*

[3]*Mössbauer Spectroscopy Laboratory, Institute of Technology, University of the National Education Commission, Podchorążych 2, 30-084 Kraków, Poland*

[4]*The Henryk Niewodniczanski Institute of Nuclear Physics, Polish Academy of Sciences, ul. Radzikowskiego 152, 31-342 Krakow, Poland*

[5]*Department of Chemistry, Purdue University, West Lafayette, IN, USA*

[6]*Chemical Consultant, Racine, WI, USA*

[7]*Univ. Grenoble Alpes, Grenoble INP, CEA, IRIG, PHELIQS, 38000 Grenoble, Franc*

[8]*AGH University of Krakow, Faculty of Geology, Geophysics, and Environmental Protection, Aleja Mickiewicza 30, 30-059 Kraków, Poland*

[9]*Institut Néel, CNRS & Univ. Grenoble Alpes, 38042 Grenoble, France*

[10]*Department of Physics, Faculty of Science, University of Zagreb, Bijenicka cesta 32, HR-10000 Zagreb, Croatia*

[*]Email: wtabis@agh.edu.pl

Seeking to unravel the enigmatic Verwey transition and its interplay with magnetism, we have conducted comprehensive measurements on the temperature-dependent electrical resistivity and magnetic moment of stoichiometric and doped-magnetite single crystals at temperatures reaching 1000 K. These investigations have allowed us to identify the Curie temperature, $T_C$, and other characteristic temperatures of the electrical resistivity. Remarkably, we have identified correlations between these temperatures and the Verwey temperature, $T_V$, indicating that the electrical transport properties and the mechanism of the Verwey transition are closely related to the magnetic properties.

Magnetite, $Fe_3O_4$, is a half metal and it stands as the oldest documented magnetic material, with a ferrimagnetic transition at $T_C$ = 853 K. The room temperature structure of magnetite is cubic, *Fd-3m*, iron atoms occupy two positions with different oxygen coordination: octahedral (B) and tetrahedral (A) with all A-sites occupied by $Fe^{3+}$ ions, resulting in an inverse spinel structure, and the B-sites occupied by $Fe^{2+}$ ($d^6$, $t_g^4$ $e_g^2$) and $Fe^3$ ($d^5$, $t_g^3$ $e_g^2$) ions. As we will show below, the cation distribution between both sites changes as the temperature is increased beyond 330 K.[1]

One of the key phenomena, still largely unexplained, is the occurrence of a metal-insulator phase transition at $T_V$ = 125 K, known as the Verwey transition[2]. Despite extensive research, with much of the evidence linking the Verwey phase transition to electron-electron and electron-phonon interactions,[3] certain aspects remain elusive. For instance, the electronic system is highly sensitive to even small disturbances, such as slight doping or deviations from stoichiometry,[4] and exposure to electromagnetic pulses of varying wavelengths,[5] which can reveal phases typically hidden in the equilibrium state. However, the most critical unresolved question is how magnetic degrees of freedom contribute to the mechanism of the Verwey transition.

Particularly, it is of high interest to understand the low-temperature atomic octahedral arrangement of the Fe atoms in cigar-like structures (so called trimerons).[6] As temperature increases above $T_V$, the electronic configuration undergoes a substantial reorganization. The *"extra"* electrons from $Fe^{2+}$ B-sites become excited (as reflected in the large peak in heat capacity)[7,8,9] and start to oscillate among all B-sites contributing to the hundredfold increase of DC conductivity. The electronic transport takes place in the minority spin band of the B-site iron. There is just one ↓ electron for every two B-sites in a $t_{2g}$ band, which is sufficiently narrow for the electrons to form magnetic polarons. Conduction is by thermally assisted hopping of these magnetic polarons from one B-site to the next. Yet, recent studies indicate that the low-temperature electronic order persists locally up to $T_C$, manifesting as short-range order (SRO) distortions,[10] thus validating $T_C$ as the onset of the spontaneous magnetization as well as of these magnetic polarons. Finally, the occurrence of a rearrangement of the cation distribution between the A and B-sites as a function of temperature, with as much as 10-15% of $Fe^{2+}$ at the A-sites at $T_C$ was documented.[1,11,12,13,14] The observed effects above $T_V$, including the cation rearrangement and its impact on SRO, are intricately linked with changes in the electric conductivity of magnetite. Specifically, these effects manifest as modifications of the mechanism of DC conductivity in the *Fd-3m* phase, as proposed by Ihle *et al.*[15] and observed experimentally[1,16,17].

Stoichiometric magnetite exhibits robust ferrimagnetism primarily governed by the dominant antiferromagnetic ($J_{AB}$ =-4.8 meV) superexchange interaction between antiparallel ferromagnetically ordered A and B Fe ions. In a first approximation, it is this antiferromagnetic interaction that leads to a rather high transition temperature, $T_C$. A-A Fe interactions are generally antiferromagnetic, with the nearest-neighbors and the next nearest-neighbors A-A exchange ($J_{NNAA}$ =-0.35 meV, $J_{NNNAA}$ =-0.2 meV) respectively, however, stronger superexchange interactions align the moments parallel[18]. The ferromagnetic B-B exchange emerges from a blend of superexchange, double exchange, and direct exchange mechanisms; though relatively weak ($J_{BB}$ =0.44 meV), it increases by 50% to reach 0.69 meV above the $T_V$. While this increase translates into only a 0.1% increase in magnetization saturation,[19] the magnetic polarons arising at $T_C$,[10] order at $T_V$ in the form of trimerons and, beyond their involvement in the $J_{BB}$ interactions, they also play a role in $J_{AB}$ interactions[20]. Consequently, magnetic polarons (and trimerons) actively contribute to both the Verwey transition and magnetism, suggesting that some of the mechanisms giving rise to the Verwey transition are prepared as early as the onset of magnetic ordering at $T_C$.

The primary objective of this study is to establish an experimental correlation between the Verwey transition temperature and the Curie temperature. To that end, we measured magnetization, AC magnetic susceptibility, and electronic resistivity in a total of 15 stoichiometric single crystals, as well as Zn, Mn, Al, and Ti-doped crystals, with dopants selected to replace Fe at the A-site, B-site, or both[7,21,22,23,24,25]. The range of doping was extensive, leading to a wide variation in $T_V$ across samples, ranging from 122.8 K for the stoichiometric crystal $Fe_3O_4$#2, to as low as 71.7 K for $Fe_{3-x}Ti_xO_4$#2, with the Verwey transition entirely suppressed in the case of sample $Fe_{3-x}Mn_xO_4$#1. Our findings demonstrate a robust correlation between the Curie temperature and the Verwey temperature.

All measurements were conducted on single crystalline samples of stoichiometric magnetite and doped with Zn, Mn, Al, and Ti, grown at Purdue University using the skull melter method (further details are provided in Supplemental Material (SM) and references 26,27,28). The only exception was one stoichiometric crystal (Fe$_3$O$_4$#1), which was grown using the same optical floating zone furnace as samples used in the previous heat capacity study[7]. Basic sample characterization results, including doping level $x$ (measured by microprobe), $T_V$ and its uncertainty (measured by AC susceptibility), $T_C$ and its uncertainty (measured by the temperature dependence of magnetic moment $\mu$), and temperatures of characteristic features of resistivity, are presented in Table 1.

TABLE I. Doping level, $x$ and characteristic temperatures ($T_V$, $T_C$, $T_{RMAX}$, $T_{RINF}$, $T_M$, $T_I$) and their uncertainties of Fe$_{3-x}$M$_x$O$_4$ (where M = Zn, Ti, Al, Mn) single crystals studied. The experimental details of the AC susceptibility, magnetic moment and resistance measurements are described in SM.

| Sample | Doping level | Verwey temperature | | Characteristic temperatures from resistivity measurements | | | | Curie temperature | |
|---|---|---|---|---|---|---|---|---|---|
| | | $T_V$ (K) | $\Delta T_V$ (K) | $T_{RMAX}$ (K) | $\Delta T_{RMAX}$ (K) | $T_{RINF}$ (K) | $\Delta T_{RINF}$ (K) | $T_C$ (K) | $\Delta T_C$ (K) |
| Fe$_3$O$_4$#1 | - | 121.5 | 0.6 | 788 | 3 | 885 | 4 | 846 | 2 |
| Fe$_3$O$_4$#2 | - | 122.8 | 0.7 | 788 | 1 | 884 | 2 | 842 | 3 |
| Fe$_{3-x}$Zn$_x$O$_4$#1 | 0.038 | 80.3 | 5.0 | 765 | 13 | 867 | 10 | 821 | 2 |
| Fe$_{3-x}$Al$_x$O$_4$#1 | 0.041 | 93.0 | 3.0 | 759 | 8 | 859 | 9 | 835 | 2 |
| Fe$_{3-x}$Mn$_x$O$_4$#1 | 0.152 | - | - | 753 | 9 | 861 | 4 | 798 | 2 |
| Fe$_{3-x}$Mn$_x$O$_4$#2 | 0.011 | 110.8 | 1.5 | 782 | 7 | 879 | 2 | 851 | 2 |
| Fe$_{3-x}$Mn$_x$O$_4$#3 | 0.011 | 109.0 | 1.5 | 777 | 1 | 881 | 2 | 842 | 1 |
| Fe$_{3-x}$Mn$_x$O$_4$#4 | 0.010 | 111.3 | 0.8 | 787 | 3 | 887 | 3 | 840 | 1 |
| Fe$_{3-x}$Ti$_x$O$_4$#1 | 0.010 | 114.5 | 0.7 | 783 | 4 | 882 | 8 | 838 | 2 |
| Fe$_{3-x}$Ti$_x$O$_4$#2 | 0.066 | 71.7 | 4.0 | - | - | 844 | 4 | 809 | 1 |
| Fe$_{3-x}$Ti$_x$O$_4$#3 | 0.004 | 119.0 | 1.0 | 767 | 13 | 876 | 1 | 835 | 2 |
| Fe$_{3-x}$Ti$_x$O$_4$#4 | 0.004 | 118.5 | 0.6 | 764 | 4 | 874 | 4 | 850 | 2 |
| Fe$_{3-x}$Ti$_x$O$_4$#5 | 0.004 | 116.0 | 2.0 | 779 | 1 | 879 | 1 | 844 | 2 |
| Fe$_{3-x}$Ti$_x$O$_4$#6 | 0.012 | 111.5 | 1.0 | 770 | 2 | 876 | 3 | 836 | 1 |
| Fe$_{3-x}$Ti$_x$O$_4$#7 | 0.013 | 110.2 | 1.0 | 770 | 10 | 864 | 8 | 848 | 2 |

The temperature dependence of magnetic moment $\mu$ (used for $T_C$ determination) and resistivity up to 1000 K are presented in Figs.1a and b, while all AC susceptibility curves, used for $T_V$ determination, are shown in Fig. S1 in SM, where also $\mu(T)$ in the full temperature range 300 K-1000 K are presented.

As presented in Fig. 1b, resistivity exhibits a local minimum at $T_M$ within the temperature range of 300–400 K. Between approximately 400 K and 700 K, resistivity displays quasi-metallic behavior, increasing with temperature. As the temperature further rises, it begins to saturate, marked by the first inflection point $T_I$ in the $\rho$ vs. $T$ curves. It subsequently decreases, resulting in resistivity maximum at $T_{RMAX}$ in the region of 750–780 K. This behavior is consistent across all studied samples. Notably, in the case of pure magnetite, this trend aligns with the findings of Elnaggar *et al.*,[1] where $T_{RMAX}$ was observed at a slightly lower temperature (840 K) in comparison to $T_C$ at 850 K. The decrease in resistivity beyond $T_{RMAX}$ suggests the onset of additional conduction channels that can be associated with a significant redistribution of charges between iron ions at tetrahedral A and octahedral B sites.[1] This redistribution could, conversely, impact both the A-B superexchange and the B-B double exchange interactions, thereby influencing the resistance mechanism, as further discussed below. Finally, resistivity begins to stabilize already above $T_C$ as evidenced by the inflection point at $T_{RINF}$. Definition of the four characteristic temperatures of the resistivity curves, $T_M$, $T_I$, $T_{RMAX}$, and $T_{RINF}$, is presented in Fig.1c, where comparison of $\mu(T)$ and $\rho(T)$ is illustrated for Mn-doped sample. It is worth noting that in the case of Mn-doped Fe$_{3-x}$Mn$_x$O$_4$#1 sample, which has the lowest Curie temperature ($T_C = 797.5 \pm 2$ K), the AC susceptibility measurements conducted down to 77 K, and magnetization measured down to 5 K did not show any signature of the Verwey transition. Additionally, Fig. 2 displays the $T_C$ and $T_V$ values relative to sample doping and nonstoichiometry parameters, derived from both our current experiments and from the literature[29].

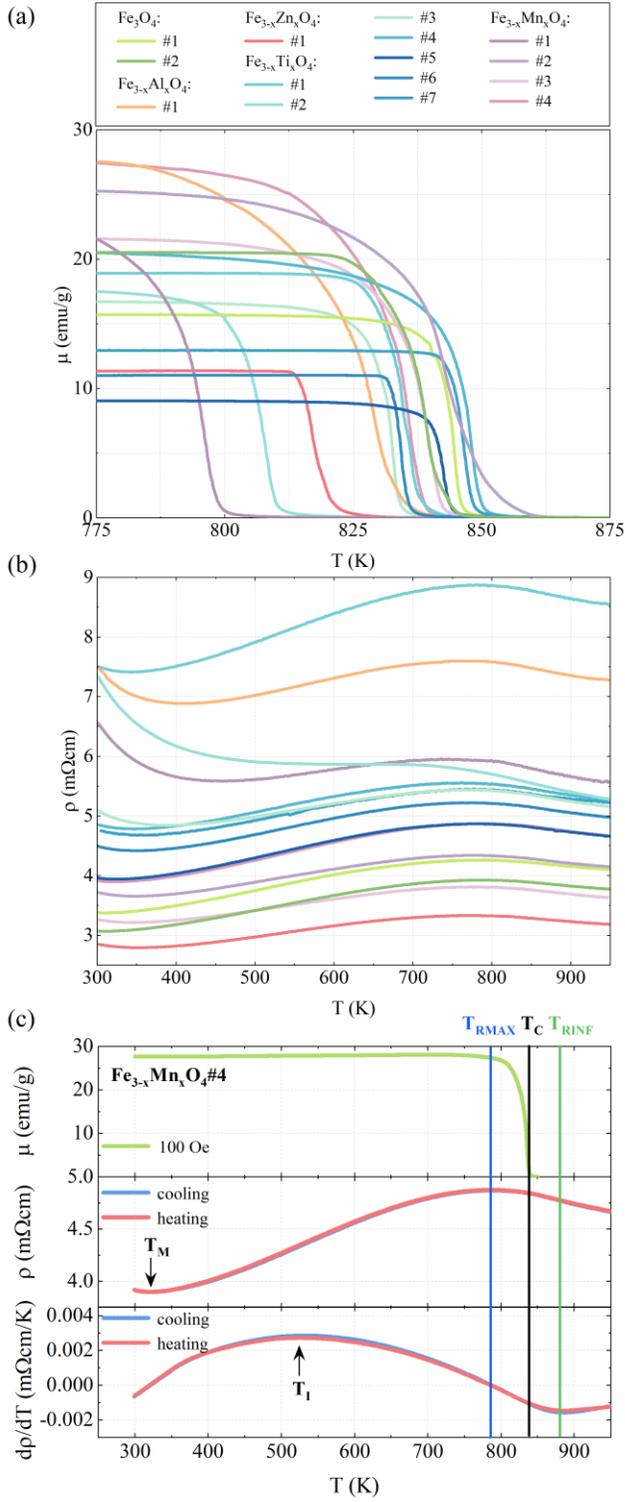

FIG. 1. Summary results from the magnetization and electronic transport measurements along with the method of the determination of the characteristic temperatures. (a) Complete set of magnetic moment vs. temperature measurements, performed at 100 Oe. (b) Resistivity as a function of temperature data collected for the same set of samples. (c) Magnetization, resistivity, and derivative of resistivity as a function of temperature of $Fe_{3-x}Mn_xO_4$#4 sample, illustrating how $T_C$ was determined from magnetic moment measurement along with the definition of characteristic temperatures extracted from resistivity (and its derivative), i.e. $T_M$, $T_I$, $T_{RMAX}$, and $T_{RINF}$. The data was collected on both cooling and heating.

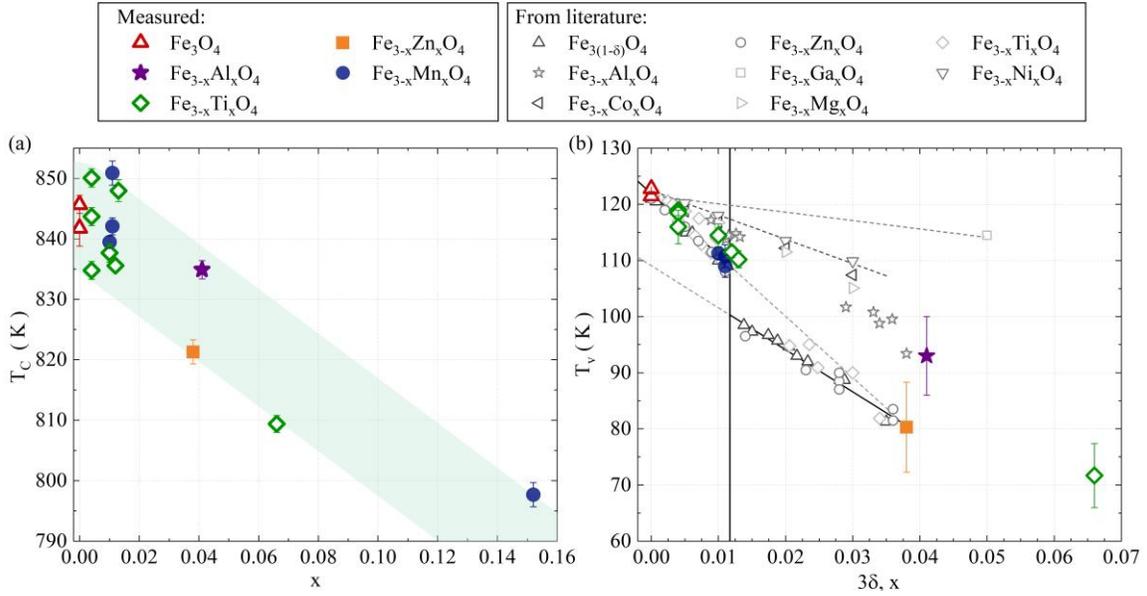

FIG. 2. Characteristic temperatures of the studied samples as a function of doping. The dependence of Curie temperature ($T_C$) and Verwey temperature ($T_V$) on doping level (x) and oxygen stoichiometry ($\delta$) are shown in panels (a) and (b), respectively. Colored symbols represent the results from this study, while gray symbols indicate data from the literature [29].

The most important results of our study are summarized in Fig. 3, where the Curie temperature and the two characteristic resistivity temperatures $T_{RMAX}$ and $T_{RINF}$ are plotted as a function of $T_V$. On decreasing $T_V$ (by increasing the doping level) $T_C$, $T_{RMAX}$, and $T_{RINF}$ also decrease. Such a correlation clearly demonstrates an interplay of electronic, lattice and magnetic degrees of freedom in magnetite. Extensive studies covering a large range of temperatures have underlined the presence of related features around $T_V$ and $T_C$, thus providing experimental evidence for the possible link between both transitions.[1,10,11,12,13,14] Note that in the case of Mn-doped $Fe_{3-x}Mn_xO_4$#1 sample, where the Verwey transition is suppressed due to the high doping level, the corresponding values of $T_{RMAX}$ = 753 K, $T_{RINF}$ = 861 K, and $T_C$ = 798 K are the lowest among the samples (refer to Table 1), further reinforcing the correlation depicted in Fig. 3. This observation, coupled with the insights from Fig. 2, suggests a consistent trend wherein the magnetic transition, the Verwey transition, and the characteristic resistivity temperatures generally decrease with an increase in dopant content. This trend holds true regardless of the specific location of the dopant atoms (Fe atoms are replaced by Zn dopants at the A sites[20], Ti and Mn in B[22,23,24], and by Al in both sites[4]), the deformation arising from the magnetic polarons, or the impact on the electronic system caused by multivalent impurities. Indeed, the impact of individual dopants can vary significantly: differences in electronic structure and resulting magnetic interactions may occur even for the same dopant, influenced by factors such as strain and defect geometry, which can depend on the growth and annealing processes[30]. This variability could explain the relatively large scattering of data observed in both $T_C$ versus doping (Fig. 2a) and, subsequently, as a function of $T_V$ in Fig. 3. However, the broad perspective provided by Fig. 3, derived from studies on a wide variety of samples, suggests that the correlation between $T_C$, resistivity features $T_{RMAX}$ and $T_{RINF}$, and $T_V$ is a general phenomenon in magnetite. A similar correlation, involving a simultaneous decrease upon Zn doping, has been reported for $T_V$ and the isotropy point $T_{IP}$,[31] where the dominant cubic magnetocrystalline anisotropy constant changes sign.

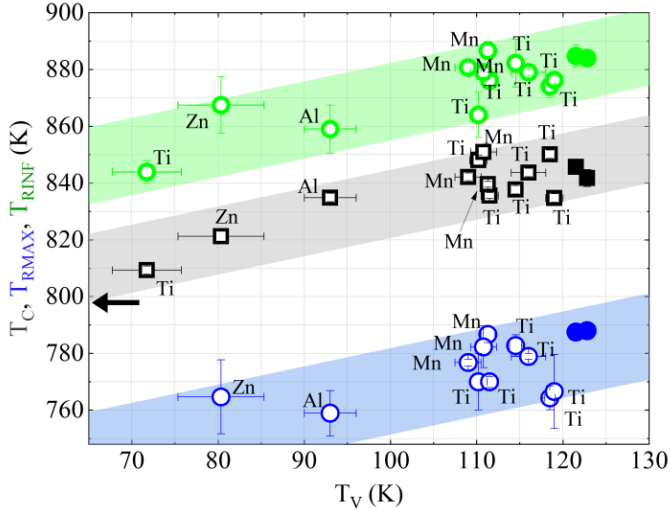

FIG. 3. Correlation of magnetic transition temperatures and resistivity features in stoichiometric and doped magnetite. The Curie temperature $T_C$ (black squares), the resistivity maximum $T_{RMAX}$ (blue circles), the upper inflection point, i.e. the minimum of the resistivity derivative $T_{RINF}$ (green circles), plotted as a function of the Verwey temperature $T_V$ for stoichiometric (solid markers) and doped (open markers) magnetite samples measured in this study. For the doped samples, the specific dopant elements are indicated next to each data point. The black arrow highlights the $T_C$ of the most heavily doped sample (Mn), for which $T_V$ is suppressed. The shaded bands are included as visual guides.

The results of certain experiments may seem to contradict the correlation between $T_C$ and $T_V$ discussed above. For example, hydrostatic pressure tends to lower $T_V$, destabilizing the low-$T$ $Cc$ crystal structure, yet it increases $T_C$ as well as $T_{IP}$ in both pure and Zn-doped magnetite.[32,33,34,35] However, interpreting the influence of external stress and resulting strain on magnetite properties is complex. While hydrostatic pressure generally lowers $T_V$, uniaxial pressure has been found to increase it[36]. Consequently, it is plausible that strain-induced coupling between electronic, lattice, and magnetic degrees of freedom influences the behaviour of all relevant temperatures, including $T_C$, $T_V$, and $T_{IP}$, especially within a narrow doping range. Thus, while we propose that magnetic interactions and the mechanism of Verwey transition are linked, coupling with intrinsic lattice strains is a factor in the observed phenomena. Lattice strains, which may be enhanced by dopants or other lattice distortions in magnetite, can disrupt the delicate balance of Coulomb forces within the crystal, leading to the ferrimagnetic order, also influencing the observed transitions. Thus, although the uncontrolled strain field renders the overall effect within a narrow doping range less clear, the general correlation between $T_C$ and $T_V$ becomes apparent when considering a broader range of dopants.

Additionally, other resistivity-related features, such as the temperature $T_M$ where the resistivity reaches a minimum, and the temperature $T_I$ corresponding to the first inflection point of the resistivity curve, both increase with decreasing $T_V$, as illustrated in Fig. 4S. However, both these resistivity features stem from a complex interplay of at least two distinct mechanisms of DC conductance in magnetite, suggesting that they may not be intrinsic indicators of electrical resistivity. Indeed, the temperature dependence of conductivity in the discussed temperature regime was studied e.g. by Ihle and Lorenz[15]. Their work successfully explained experimental data for stoichiometric magnetite up to 600 K by considering both the short-range electronic order at $T<T_V$ and $T>T_V$ (in the latter case the local ordering resembles the $Cc$ symmetry) terminating at $T_C$[10,37], and small polaron correlations. Two primary mechanisms of conduction were considered: small polaron-band conduction, which dominates at lower temperatures (and persists above $T_V$), and small polaron hopping, which is strongly influenced by the SRO. Our two stoichiometric samples exhibit resistivity behavior consistent with this model, though slight variations due to differences in growth conditions and annealing processes are expected (see Fig. 3Sb of SI, where the resistivity data from one of our samples[1], along with other literature data for stoichiometric magnetite, is compared with the theoretical predictions). However, dopants, which occupy different positions within the magnetite lattice can easily disrupt the balance between both mechanisms, causing shifts in $T_M$ and $T_I$ along different directions. On the contrary, we believe that high-temperature resistivity features such as $T_{RMAX}$ and $T_{RINF}$ have a more intrinsic character since in the model proposed by Ihle *et al.*, at such elevated temperatures, the hopping mechanism predominantly governs the temperature dependence of conductivity.

In conclusion, through extensive measurements of $T_V$, $T_C$, and the temperature-dependent resistivity in numerous stoichiometric and differently doped (element and lattice site) magnetite single crystals, we have experimentally established a correlation between the Verwey transition temperature, Curie temperature, and other characteristic temperatures extracted from electronic transport experiments above 300 K. This correlation is striking considering the considerable disparity of up to 700 K between $T_V$ and $T_C$. Our findings strongly suggest an intricate relationship between the Verwey transition and magnetic interactions, pointing towards a unified mechanism governing both high and low-temperature behaviors of magnetite. This insight not only sheds light on the fundamental physics of magnetite but also underscores the critical role of magnetism in driving the complex behaviors observed in this material.

## ACKNOWLEDGMENTS

The work was supported by the National Science Centre, Poland, Grant No. OPUS: UMO-2021/41/B/ST3/03454, the Polish National Agency for Academic Exchange under "Polish Returns 2019" Programme: PPN/PPO/2019/1/00014, and the subsidy of the Ministry of Science and Higher Education of Poland. D.R. acknowledges financial support of National Science Centre, Poland (Grant No. 2018/30/E/ST3/00377). EPMA measurements were financially supported by AGH University of Krakow from the research subsidy 16.16.140.315 (A.W.).

**Supplemental Material for:**
**Correlation between magnetism and the Verwey transition in magnetite**

**Experimental techniques**

AC susceptibility measurements were conducted using a custom-built magnetometer featuring a standard counter-wound coil set. A chromel vs. Au-0.03 at.% Fe thermocouple by Lake Shore, positioned near the sample, precisely recorded the sample's temperature. Measurements were performed continuously, with a temperature ramp rate of no more than 0.2 K/min. Sample response was detected using an SR830 lock-in nanovoltmeter. Typical measurement parameters included a voltage of 5 V supplied to the transmitting coil, a frequency of 188.88 Hz, a time constant of 300 ms, and a sensitivity of 20–50 mV, depending on the sample's mass. The results from AC susceptibility measurements, crucial for determining the $T_V$, are presented in Fig. 1Sa. $T_V$ is defined as half of the change in the susceptibility signal.

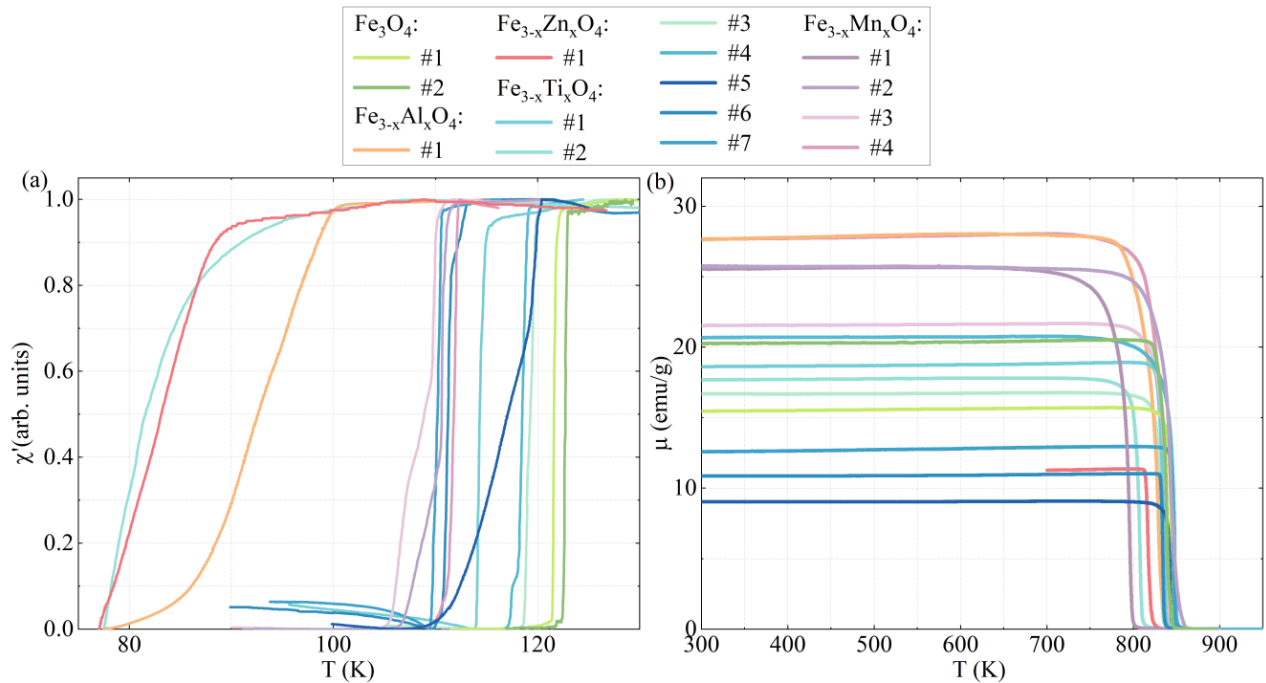

Fig.1S. AC magnetic susceptibility and magnetic moment $\mu$ vs. $T$ (in whole temperature range) results for all measured samples are presented in panel a) and b), respectively.

AC susceptibility measurements were also employed to assess the impact of the high temperatures experienced by the samples during resistance and magnetic moment measurements on the Verwey transition. The value of $T_V$ was verified at each stage of the measurements. Interestingly, it was noted that subjecting the sample to high temperatures did not alter $T_V$ significantly. However, for certain individual samples, a slight widening of the Verwey transition was observed.

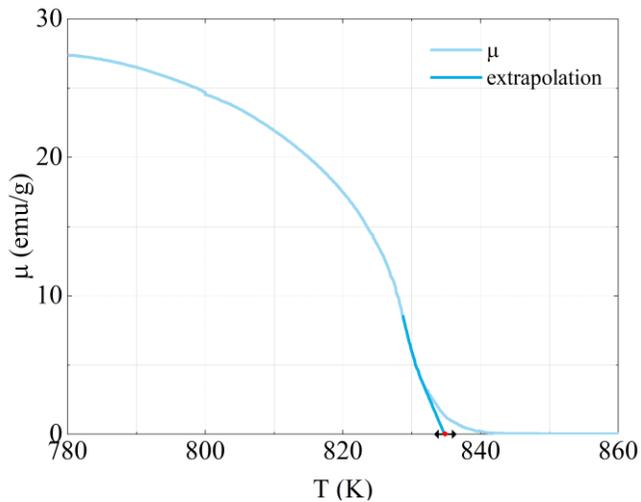

FIG. S2. Graphical illustration of the linear extrapolation method used to determine the $T_C$ value for sample $Fe_{3-x}Al_xO_4$#1. The light blue color corresponds to the measured data, while the dark blue line is the extrapolated line. The uncertainty in determining of $T_C$ was assumed to be a reasonable temperature range in which the magnetization curve could drop to zero if a different range of points was considered in the extrapolation.

The Curie temperature for each sample was determined through the temperature dependence of the magnetic moment. Magnetic measurements were conducted for all samples in the temperature range from 300 K to approximately 950 K, employing a VSM magnetometer option in the Quantum Design PPMS system. To ensure stable conditions during measurement, the sample, placed on the holder, was coated with a specialized cement (Zirac Cement QDS) to prevent displacement and ensure proper thermalization. Additionally, the sample holder included a heater and a thermometer. The sample, along with the heating element and thermometer, was encased in copper foil, serving as a screen to ensure uniform temperature distribution across the sample area. This prepared sample was then positioned in the measuring setup. After multiple testing runs, it was determined that a magnetic field strength of 100 Oe provided a sufficiently strong signal without significantly widening the magnetic transition. The temperature dependence of the magnetic moment for each of the samples are presented in Fig 1Sb. Furthermore, to precisely ascertain the temperature at which magnetization ceases (Curie temperature), each curve was extrapolated within the low magnetization values, as illustrated in Fig. 2S. Subsequently, based on this extrapolation, the $T_C$ value of each sample was determined.

Four-point resistivity measurements were conducted using an SR-830 lock-in amplifier. Four copper wires were securely attached to the sample surface using high-temperature silver paste DuPont 6838. The resistivity measurements were performed over the temperature range of 300 to 973 K, with a ramp rate of 3 K/min, in an atmosphere of technical-grade argon with a purity of 99.999%. The excitation current, with a frequency of 73.33 Hz, was provided by the SR-830 Lock-in nanovoltmeter with a 5V output and a 1000 $\Omega$ shunt resistor, connected in series with the sample. This configuration resulted in a circuit current of 4.74 mA, as directly measured with a Fluke amperemeter. The voltage drop across the sample was then measured directly at the input of the nanovoltmeter. Considering the uncertainties associated with sample and contact geometry, the total uncertainty in resistivity ($\rho$) for different samples is estimated to reach a maximum of 10%.

**Sample characterization**
Single crystalline magnetite stoichiometric ($Fe_3O_4$#2) and Zn, Ti, Al and Mn doped samples were grown from the melt by the cold crucible technique (skull melter)[1], at Purdue University, USA from at least 99.99% pure $Fe_2O_3$ material. Subsequently, the crystals underwent subsolidus annealing under $CO/CO_2$ gas mixtures to achieve the desired metal/oxygen ratio.[2,3,4] The $Fe_3O_4$#1 stoichiometric magnetite single crystal was grown by the floating zone method. The composition and uniformity of each sample were confirmed using a microprobe electron analyzer.

In case of low concentration, dopants substitute Fe with different valence and in different places:

Zn: $\equiv Fe_{3-x}Zn_xO_4 \equiv (Fe^{3+}_{1-x}Zn^{2+}_x)[Fe^{3+}_{1+x}Fe^{2+}_{1-x}]O_4$

Mn: $\equiv Fe_{3-x}Mn_xO_4 \equiv (Fe^{3+})[Fe^{3+}_1 Fe^{2+}_{1-x}Mn^{2+}_x]O_4$

Ti: $\equiv Fe_{3-x}Ti_xO_4 \equiv (Fe^{3+})[Fe^{3+}_{1-2x}Fe^{2+}_{1+x}Ti^{4+}_x]O_4$

Al: $\equiv Fe_{3-x}Al_xO_4 \equiv (Fe^{3+}_{1-0.15x}Al^{3+}_{0.15x})[Fe^{3+}_{1-0.85x}Fe^{2+}Al^{3+}_{0.85x}]O_4$

Here, () brackets denote tetrahedral (A) positions, while [] brackets denote octahedral (B) positions. The valence of Mn was assumed to be +2, as documented in Ref. 5, 6, and for Al, this is considered the most probable alternative[4]. Due to their different lattice positions, dopants have a minimal impact on the lattice, electronic structure, and magnetic moment, except perhaps for $Fe_{3-x}Mn_xO_4$#1, where their influence may be more pronounced. For instance, whereas $Zn^{2+}$ (substituting Fe on A sites) and $Ti^{4+}$ (occupying B sites) cations expand the lattice, $Al^{3+}$ (inhabiting both positions) contracts it[7].

Verwey transition temperatures were drawn from magnetic susceptibility vs. T measurements; the results are shown in Fig. 1Sa, while magnetic moment $\mu(T)$ in the full measured range 300 K-1000 K is in Fig. 1Sb.

### Resistance vs. magnetism

In Fig. 3Sa, the resistivity vs. temperature results for stoichiometric magnetite[8] are depicted, overlaid with the experimental values from Ref. 9, and the outcomes from a theoretical small polaron model[10]. It's worth noting that all samples analyzed in this paper underwent measurements utilizing the same equipment as the stoichiometric magnetite sample[8], employing a similar protocol.

Furthermore, $T_{RMAX}$ and $T_{RINF}$ vs. $T_C$ drawn from Fig. 3 of the main text showing their quasilinear correlation are presented in Fig. 3Sb.

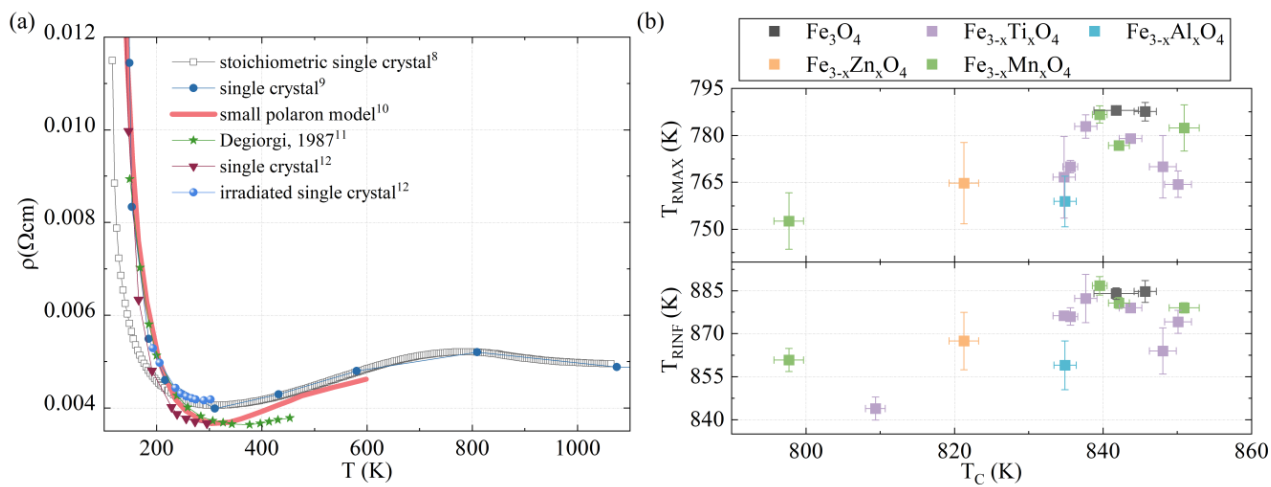

FIG. S3. Resistivity in magnetite. (a) Selected results for temperature dependence of resistivity of magnetite. The results for stoichiometric magnetite cited after Ref 8, were collected on one of our stoichiometric samples using the same equipment as for the resistivity measurements on the samples presented in this paper. Note that two results cited after Ref. 12, concerns a pristine single crystal, and the irradiated one, showing that defects substantially affect resistivity, as also observed here. (b) Quasilinear $T_{RMAX}$ and $T_{RINF}$ vs. $T_C$ correlation drawn from Fig. 3 of the main text.

Finally, the relationship between other characteristic temperatures derived from the resistivity versus temperature measurements, specifically $T_M$ and $T_I$, with respect to $T_V$ is illustrated in Fig. 4S(a) and (b). Both $T_M$ and $T_I$ show an inverse correlation with $T_V$.

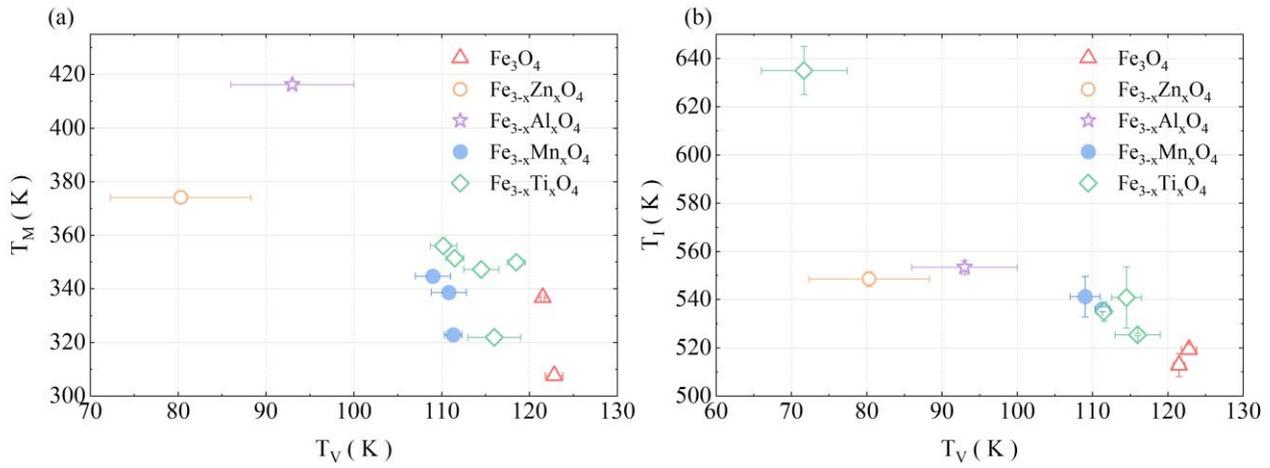

FIG. S4. Relationship between the characteristic temperatures in magnetite: (a) the resistivity minimum $T_M$ and (b) the first inflection point of the resistivity curve $T_I$ and the Verwey temperature $T_V$ for stoichiometric and doped Zn, Mn, Al, Ti magnetite samples.

**References to SM**